\documentclass[onecolumn]{IEEEtran}
\makeatletter
\def\ps@headings{%
\def\@oddhead{\mbox{}\scriptsize\rightmark \hfil \thepage}%
\def\@evenhead{\scriptsize\thepage \hfil \leftmark\mbox{}}%
\def\@oddfoot{}%
\def\@evenfoot{}}
\makeatother
\usepackage{epsfig} \usepackage{amsmath} \usepackage{amssymb}
\usepackage{amsfonts} \usepackage{graphicx} \usepackage{subfigure}
\newtheorem{theorem}{Theorem} \newtheorem{lemma}[theorem]{Lemma}

\newtheorem{corollary}[theorem]{Corollary}
\begin{document}
\title{Tunable Locally-Optimal Geographical Forwarding in Wireless
  Sensor Networks with\\ Sleep-Wake Cycling Nodes}
\author{K.P. Naveen and Anurag Kumar\\
  {Dept. of E.C.E.,}
  {Indian Institute of Science, Bangalore 560 012, India.}\\
  {\{naveenkp, anurag\}@ece.iisc.ernet.in}}
\maketitle
\begin{abstract}
  We consider a wireless sensor network whose main function is to
  detect certain infrequent alarm events, and to forward alarm packets
  to a base station, using geographical forwarding. The nodes know
  their locations, and they sleep-wake cycle, waking up periodically
  but not synchronously. In this situation, when a node has a packet
  to forward to the sink, there is a trade-off between how long this
  node waits for a suitable neighbor to wake up and the progress the
  packet makes towards the sink once it is forwarded to this
  neighbor.  Hence, in choosing a relay node, we consider the problem
  of minimizing average delay subject to a constraint on the average progress. By
  constraint relaxation, we formulate
  this next hop relay selection problem as a Markov decision process
  (MDP). The exact optimal solution (BF (Best Forward)) can be found,
  but is computationally intensive.  Next, we consider a
  mathematically simplified model for which the optimal policy (SF (Simplified Forward)) turns out to
  be a simple one-step-look-ahead rule.  Simulations show that SF is
  very close in performance to BF, even for reasonably small node
  density.  We then study the end-to-end performance  of
  SF in comparison with two extremal policies: Max Forward (MF) and First Forward (FF), and an end-to-end
  delay minimising policy proposed by Kim et al. \cite{kim-etal09optimal-anycast}. We find that, with appropriate
  choice of one hop average progress constraint, SF can be \emph{tuned} to provide a favorable trade-off
  between end-to-end packet delay and the number of hops in the
  forwarding path.
\end{abstract}
\section{Introduction}
An important application of wireless sensor networks (WSN) is dense
embedded sensing for the purpose of detecting certain infrequently
occuring events, such as failures in a large structure, or intrusion
into a secure region. Such an event can occur anywhere in a large WSN,
and once an event is detected, the alarm needs to be rapidly sent to
the sink  for further action. In such WSNs, typically the
nodes rely on batteries, or energy harvested from their surroundings,
and, hence, need to be extremely parsimonius in their use of energy.
In order to conserve energy, the nodes operate in sleep-wake cycles;
when a node wakes up it performs sensing, and also can assist in
forwarding any alarm packets towards the sink. In this paper, we
consider the situation in which the sleep-wake cycles of nodes are
\emph{not synchronized.}  In such a setting, stateful routing is not
possible. 
Instead, if the nodes
know their own locations and that of the sink, then it is possible to
dynamically select forwarding nodes that are successively nearer to
the sink.  This is called \emph{geographical routing,} and has been
widely studied as a simple scalable approach for routing in sensor
networks \cite{takagi-kleinrock84optimaltransmission,hou-etal86rangecontrol,karp-etal00gpsr,kuhn-etal08algorithmicapproach}.
 For the purpose of location determination, low cost GPS
devices are now becoming available, and can be incorporated in the
nodes; alternatively, approximate localization algorithms based on
various geometrical principles can also be used (see, for example,
\cite{dulman-etal06hop-count,nath-kumar08distance-hop}. For a
survey on routing and localization, see
\cite{mauve-hartenstein01survey,Akkaya-younis05survey}. 
   In this paper we assume that nodes know their exact
locations and also the location of the sink.

\noindent
\emph{The relay node selection problem:} In geographical forwarding,
in our setting, there arises the problem of optimal relay node
selection, which we now discuss.  One approach is that of \emph{greedy
  forwarding}, in which an intermediate node forwards the packet to
its neighbor node that makes maximum progress towards the sink. This
scheme is referred to as Most Forward within Radius (MFR)
(\cite{takagi-kleinrock84optimaltransmission,hou-etal86rangecontrol}).
If the node density is large such that every node has a neighbor that
is closer to the sink than itself, then the greedy approach can find
routes close to the minimum hop paths. Following a minimum hop path is
beneficial since it reduces the number of times the network needs to
transmit the packet.

However, when the nodes are sleep-wake cycling in an asynchronous
manner there is a trade-off between the delay in relay node selection
and the progress made towards the sink. For example, if MFR is
implemented, then for an intermediate node to forward the packet to a
relay node that makes the maximum progress towards the sink, the
intermediate node will need to wait for all its neighbors closer to
the sink than itself to wake up. This will result in an increase in
the delay of the alarm that is being forwarded. In fact, a counterpart
to the MFR policy could be the policy that forwards the packet to the
first node that wakes up and is nearer to the sink than the
intermediate node. In this paper we call this latter policy First
Forward (FF), and the MFR policy, simply, Max Forward (MF).

In this paper we study the above trade-off for the following one hop
relaying problem. A node needs to forward a packet to the sink. There
is a set of neighbors of the node that are nearer to the sink than the
node; the \emph{forwarding set}. The nodes are asynchronously
sleep-wake cycling according to a certain model. We seek policies for
relay node selection so as to minimize the delay in determining the
relay node, subject to a constraint on the progress made towards the
sink. We assume that each node has at least one neighbor that is strictly
closer to the sink than itself so that greedy forwarding will always
find a path to sink.  This is a reasonable assumption for large node
densities.

\noindent
\emph{Our contributions:} 
\begin{itemize}
\item The problem of minimizing average one hop delay subject to a
  constraint on the average progress made, when nodes wake up
  periodically, but not synchronously, is formulated as a Markov
  decision problem (MDP), and solved to yield the optimal policy which
  we call Best Forward (BF).  See Section~\ref{problem_formulation}
  and Section~\ref{optimal_policy}.
\item In a mathematically simplified setting (i.i.d., exponentially
  distributed inter-wakeup times) the MDP approach is used to derive a
  threshold type policy, called Simplified Forward (SF). The threshold
  is a function of the constraint on progress, and the policy is to
  transmit to the first node which wakes up and makes a progress of
  more than the threshold. See Section~\ref{suboptimal_policy}.  While
  such a policy has been proposed heuristically in previous works (\cite{paruchuri-etal04RAW,liu-etal07CMAC}), we 
  have derived it from the MDP formulation and we
  show through simulations that the performance of this policy is
  close to that of BF.  The simulation results are in
  Section~\ref{simulation_results}.
\item Finally, we compare the end to end performance (average delay
  and hop counts) of the SF policy with the forwarding policy proposed
  by Kim et al.\ \cite{kim-etal09optimal-anycast}. The approach of Kim
  et al.  aims to achieve minimum average end-to-end delay, but at the
  expense of an initial configuration phase. The SF policy, however,
  does not need any global organization phase, and the progress
  constraint  can be used to \emph{tune}
  the end-to-end performance to suitably trade-off between end-to-end
  delay and the number of hops in the forwarding path. These results
  are reported in Section~\ref{simulation_results}.
\end{itemize}

\section{Related Work} 
\label{related_work} 
Zorzi and Rao (\cite{Zorzi-rao03geographicrandom}) consider a scenario
similar to ours: geographical forwarding in a wireless mesh network in
which the nodes know their locations, and are sleep-wake cycling. They
propose GeRaF (Geographical Random Forwarding), a distributed relaying
algorithm, whose objective is to carry a packet to its destination in
as few hops as possible, by making as large progress as possible at
each relaying stage. Thus, the objective is similar to the MFR
algorithm, mentioned above
(\cite{takagi-kleinrock84optimaltransmission,hou-etal86rangecontrol}).  For their algorithm, the authors
obtain the average number of hops (for given source-sink distance) as
a function of the node density. These authors do not consider the trade-off between relay selection
delay and the progress towards the sink, which is a major contribution
of our work.

Liu et al.\ (\cite{liu-etal07CMAC}) propose a relay selection approach
as a part of CMAC, a protocol for geographical packet forwarding. With
respect to the fixed sink, a node $i$ has a forwarding set consisting
of all nodes that make progress greater than $r_0$ (an algorithm
parameter). If $Y$ represent the delay until the first wake-up instant
of a node in the forwarding set, and $X$ is the corresponding progress
made, then, under CMAC, node $i$ chooses an $r_0$ that minimizes the
expected normalized latency $\mathbb{E}[\frac{Y}{X}]$. The Random
Asynchronous Wakeup (RAW) protocol (\cite{paruchuri-etal04RAW}) also
considers transmitting to the first node to wake up that makes a
progress greater than a threshold $Th$.  Interestingly, this is also the structure
of the optimal policy provided by one of our Markov decision process
formulations.

Kim et al.\ (\cite{kim-etal09optimal-anycast}) consider a dense WSN in
which the traffic model and sleep-wake cycling are similar to ours. An
occasional alarm packet needs to be sent, from wherever in the network
it is generated, to the sink. The nodes are asynchronously sleep-wake
cycling. The authors develop an optimal anycast scheme to minimize
average end-to-end delay from any node $i$ to the sink. The
optimization is also done over sleep-wake cycling patterns and rates.
A dynamic programming approach is taken, with the stages being the
number of hops to the sink. While the framework is similar to ours,
Kim et al.\ do not consider the objective of spatial progress at each
hop, which results in the reduction of hop counts along the forwarding
paths, and thus in the reduction of node energy utilization. In our
work, we have studied the trade-off, at a typical forwarding stage,
between forwarding delay and the distance that the packet covers in
the hop.

Rossi et al.\ (\cite{rossi-etal08SARA}) consider the problem of
geographical forwarding in a wireless sensor network in which each node
knows its hop distance from the sink. For each link, there is a link
cost (for example, energy cost) for forwarding a packet over that
link. Thus, there are two end-to-end cost criteria for a forwarding
path: the total link cost of the path, and the number of hops in the
path. When a node, say $i$, has a packet to forward to the sink, it
has to consider the trade-off between cost reduction and hop distance
reduction; note that cost can be reduced by forwarding the packet to a
neighbor node with the same hop distance to the sink, but using which
the total link cost could be lower.  The information available at $i$
is the cost to all its neighbors, and the statistics of the
costs-to-go from the neighbors. The major difference in our work is
that we have a sequential decision problem \emph{at each stage}, since
the costs (wake-up delay) and rewards (progress towards the sink) are
revealed as the nodes wake up, and only the statistics are known a
priori.

Chaporkar and Proutiere (\cite{chaporkar-proutiere08joint-probing})
consider the problem of a transmitter that needs to transmit over one
of several available channels. The transmitter can probe the channels
to determine channel state information in order to encode its
transmissions. The trade-off is between the time taken to probe and
the throughput advantage of finding a good channel. Some important
differences between their model and ours are the following.  In our
work the trade-off is between the time taken to wait for a relay to
wake up, and the \emph{spatial} progress the relay makes towards the
sink. In \cite{chaporkar-proutiere08joint-probing}, the transmitter
can use an unprobed channel, whereas in our problem a relay that has
not yet woken up cannot be used. In
\cite{chaporkar-proutiere08joint-probing}, the transmitter can probe
the channels in an order that it can choose (e.g., the stochastically
best channel first); in our problem the relays wake up in a random
order that is not under the control of the transmitter. In
\cite{chaporkar-proutiere08joint-probing} it is shown that if the use
of an unprobed channel is not allowed then a one-step-look-ahead rule
is optimal. This is similar to the solution we obtain for a simplified
version of our model. Note that whereas the concern in
\cite{chaporkar-proutiere08joint-probing} is only with one-step
relaying, we also study how the one-step policy performs in terms of
end-to-end objectives, namely, path delay and path hop count.

\section{System Model}
\label{system_model}
\subsection{Node Deployment}
$N$ identical sensor nodes are uniformly deployed in the square region
$[0,L]^2$. We take $N$ to be a Poisson random variable of rate
$\lambda L^2$ where $\lambda$ is the node density. Let $x_i$,
$i=1,2,...,N$, be the locations of the nodes. Additional \emph{source} and
\emph{sink} nodes are placed at fixed locations $x_0=(0,0)$ and
$x_{N+1}=(L,L)$ respectively. Thus including the source and sink
nodes, there are a total of $N+2$ nodes in the disk. $r_c$ is the
communication range of each node. Two nodes $i$ and $j$ are called
neighbors if and only if $|x_i-x_j|\le r_c$. The distance between
node $i$ and sink ($N+1$) is $L_i=|x_{N+1}-x_i|$.
\subsection{The Sleep-Wake Process}
To conserve energy, each node performs periodic sleep-wake cycling.
The sleep-wake times of the nodes are \emph{not synchronized}. Since
we are interested in studying the delay incurred in routing due to
sleep-wake cycling alone, we neglect the transmission delay,
propagation delay and other overhead delays. This means that if node
$i$ has a packet to transmit to its neighboring node $j$, then $i$
can transmit immediately at the instant $j$ wakes up.  We model this
by taking the time for which a node stays awake to be zero.

More formally, let $T_i$, $i=1,2,...,N+1$ be \emph{i.i.d.} random variables
which are uniform on $[0,T]$, where $T$ is the period of the sleep
wake cycle. Then node $i$ wakes up at the periodic instants $kT+T_i,
k\ge0$.  We define the \emph{waiting time for i} to wake up at time
$t$ as,
\begin{equation}
\label{waiting_eqn}
W_{i}(t)=\inf\{kT+T_i\ge t:k\ge0\}-t
\end{equation} 
\subsection{Forwarding Rules and Assumptions}
Forwarding rules dictate the actions a node can take when it has to
transmit. We are interested in decentralized policies where a node can
take decisions only by observing the activities in its neighborhood
(\emph{i.e.,} the disk of radius $r_c$ centered around the node of
interest). In this regard we impose some restrictions on the network.

\noindent
\emph{Traffic Model:} There is a single packet in the network which is
to be routed from the source to sink.  At time $0$, the packet is
given to the source and the routing process begins. The nodes which
get the packet for forwarding are called relay nodes. The packet
traverses a sequence of relay nodes to eventually reach the sink, at
which time the routing ends.  Thus there is a single flow and further
the flow consists of only one packet. This set up is reasonable,
because in sensor networks we can assume that the events are
sufficiently separated in time and/or location so that the flows due
to two events do not intersect. To avoid multiple packet transmission
by different nodes detecting the same event, the nodes can resolve
among themselves to select one node (say the one closest to the sink),
which can then transmit. Further, the information about an event
comprises its location, and possibly target classification data, which
along with some control bits can be easily incorporated in a single
packet. This justifies the idea to study the performance of a single
packet alone. 

\noindent
\emph{Forwarding Set:} Each node knows its location and the location
of the sink. The \emph{forwarding set} of a node is the set of its
neighbors that are closer to the sink then itself.  A relay node
considers forwarding the packet only to a node in its forwarding set.
Each node knows the number of neighbors in its forwarding set, but is
not aware of their locations and wake times.  While in this paper we
assume that each node knows the number of nodes in its forwarding set,
it would be desirable to develop forwarding algorithms that do not
require even this knowledge. We leave this as future work, but in
Section \ref{end_subsec} we provide simulation results on the
performance of our algorithm when the node takes the number of nodes
in its forwarding region to be just the expected number of nodes.
\subsection{Some Notation}
\label{notation_sec}
\begin{figure}[ht]
\centering
\includegraphics[scale=0.5]{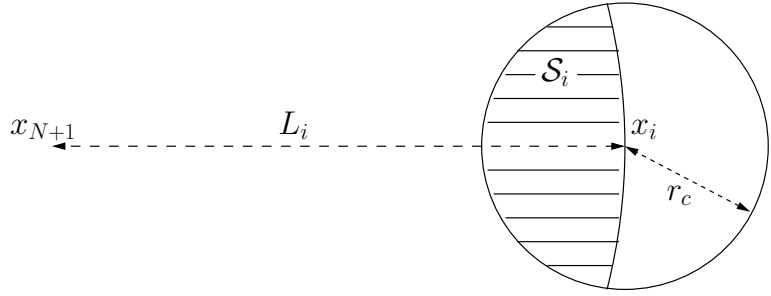}
\caption{$x_i$ and $x_{N+1}$ are the locations of node $i$ and sink
  respectively. $L_i$ is the distance between them, $r_c$ is the
  communication radius.  $\mathcal{S}_i$ is the set of all points that
  are within the communication radius of node $i$ and closer to sink
  than $i$. $\mathcal{S}_i$ is the shaded region in the figure.
  \label{forwardset_figu}}
\end{figure}

To define a forwarding policy more formally, we begin by setting up
some notation. Consider a generic node $i$ which gets the packet to
forward at some instant $t$. Let $\mathcal{S}_i=\{y:|y-x_i|\le r_c,
|x_{N+1}-y|<L_i\}$.  $\mathcal{S}_i$ is the set of all points that are
within the communication radius of $i$ and are strictly closer to the
sink than $i$ (see Fig. \ref{forwardset_figu}) (we ignore edge effects
by assuming that $\mathcal{S}_i\subset[0,L]^2$).  If
$x_j\in\mathcal{S}_i$ then the progress made by $j$ is $Z_j=L_i-L_j$.
Let $N_i$ be the number of nodes in $\mathcal{S}_i$. Note that
$N_i\sim Poisson(\lambda |\mathcal{S}_i|)$, where $|\mathcal{S}_i|$ is
the area of the region $\mathcal{S}_i$. Recall that node $i$ knows
$N_i$ and hence we focus on the event $\{N_i=K\}$ for some $K>0$.

Let the indices of the nodes in $\mathcal{S}_i$ be arranged as
$i_1,...,i_{K}$, such that $W_{i_1}(t)\le W_{i_2}(t)\le,...,\le
W_{i_{K}}(t)$.  The corresponding values of progress are
$Z_{i_1},Z_{i_2},...,Z_{i_{K}}$. For simplicity, from here on we
neglect $i$ in the subscript and simply use $W_{1}(t),...,W_{{K}}(t)$
and $Z_{1},Z_{2},...,Z_{{K}}$.

The locations of each of these $K$ nodes are uniformly distributed in
the region $\mathcal{S}_i$ independent of the others. Hence the
progress made by them are \emph{i.i.d.} whose distribution is same as $Z$.
The \emph{p.d.f.} of $Z$ is supported on $[0,r_c]$ and is given by,
\begin{equation}
\label{distribution_equn}
f_{Z}(z)=\frac{2(L_i-z)\cos^{-1}\left(\frac{{L_i}^2+{(L_i-z)}^2-{r_c}^2}{2L_i(L_i-z)}\right)}{|\mathcal{S}_i|}
\end{equation}
Where $|\mathcal{S}_i|$ denotes the area of the region $\mathcal{S}_i$, 
\begin{equation}
\label{area_equn}
|\mathcal{S}_i|=\int_0^{r_c}{2(L_i-z)\cos^{-1}\left(\frac{{L_i}^2+{(L_i-z)}^2-{r_c}^2}{2L_i(L_i-z)}\right)}dz
\end{equation}

\begin{figure}[ht]
\centering
\includegraphics[scale=0.5]{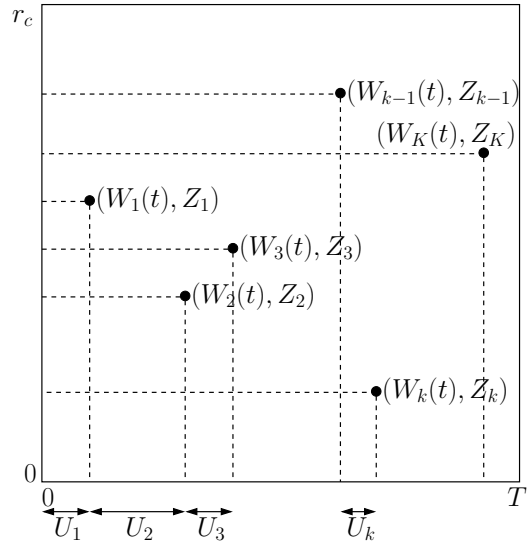}
\caption{$(W_{k}(t),Z_{k})$ represents the wake instant and the
  progress respectively, made by the Node ${i_k}$ in $\mathcal{S}_i$.
  These are shown as points in $[0,T]\times[0,r_c]$. $U_k$ is the
  inter-wakeup time between node $i_k$ and
  $i_{k-1}$.\label{wakeinstants_figu}}
\end{figure}

Let $U_1=W_{1}(t)$ and $U_k=W_{{k}}(t)-W_{{k-1}}(t)$ for $2\le k\le
K$. We refer to $\{U_k\}$ as the \emph{inter-wakeup} times. These are
the waiting times between the wakeup instants of sucessive nodes in
$\mathcal{S}_i$ (see Fig. \ref{wakeinstants_figu}). Further $U_k$ and $Z_{k}$ are independent.
 
The waiting times $W_{1}(t),W_{2}(t),$ $...,W_{{K}}(t)$ are the order
statistics of $K$ \emph{i.i.d.} random variables that are uniform on $[0,T]$.
The \emph{p.d.f.} of the $k-{th}$ order statistics is \cite[Chapter
2]{orderstatistics},
\begin{equation}
\label{orderstat_equn}
	f_{W_{k}}(u)=\frac{K! u^{k-1}(T-u)^{K-k}}{(k-1)!(K-k)!T^K}
\end{equation}
for $0\le u\le T$.  Also the joint \emph{p.d.f.} of the $k-{th}$ and $l-{th}$
order statistics (for $k<l$) is \cite[Chapter 2]{orderstatistics},
\begin{equation}
\label{jointorderstat_equn}
	f_{W_{k},W_{l}}(u,v)= \frac{K!u^{k-1}(v-u)^{l-k-1}(T-v)^{K-l}}{(k-1)!(l-k-1)!(K-l)!T^K}
\end{equation}
for $0\le u<v\le T$. Later we will be interested in  the  conditional \emph{p.d.f.}  $f_{U_{k+1}|W_{{k}}}$ for $1\le k\le K-1$. Using the above equations we can write ,
\begin{eqnarray}
	\label{condorderstat_equn}
	f_{U_{k+1}|W_{{k}}}(u|w)&=&\frac{f_{W_{{k}},W_{{k+1}}}(w,w+u)}{f_{W_{{k}}}(w)}\nonumber\\
	&=&(K-k)\frac{(T-w-u)^{K-k-1}}{(T-w)^{K-k}}
\end{eqnarray}
for $0\le w\le T$ and $0\le u\le T-w$.

\subsection{Single Hop Policy}\label{policy_sec}
Decision process begins at the instant $t$ at which node $i$
gets the packet to forward. This is \emph{stage} $k=0$.  The $k-th$
($k\ge1$) decision instant is the time at which node $i_k$ wakes up.\\
A \emph{\textbf{Single Hop (SH) policy $\pi$}} is a sequence of
mappings $\{\mu_k^{\pi}:0\le k\le K\}$, where
$\mu_0^{\pi}:\{(0,0)\}\rightarrow \{0\}$ and for $ k\ge 1$
$\mu_k^{\pi}:{[0,T]\times[0,r_c]}\rightarrow \{0,1\}$. $\pi$ should
also satisfy $\mu_{K}^{\pi}(w,b)=1$. The function $\mu_{k}^{\pi}$
maps the state at stage $k$ to an action $0$ (continue) or $1$ (stop).  Let $D^{\pi}(t)$ and
$Z^{\pi}(t)$ denote the delay incurred and progress made by node $i$
using policy $\pi$.  Forwarding rules for node $i$, using policy $\pi$
are as follows:
\begin{itemize}
\item At stage $0$, node $i$ has to wait for further nodes to wake up.
  We represent this by allowing the only \emph{state} at stage $0$ to
  be $\textbf{0}=(0,0)$ and the corresponding action to be to $0$
  (continue to wait) \emph{i.e.,} $\mu^{\pi}_0(\textbf{0})=0$.
\item If $L_i\le r_c$, then wait for sink to wake up and transmit to
  it.  In this case, the delay and progress made are
  $D^{\pi}(t)=W_{N+1}(t)$ and $Z^{\pi}(t)=L_i$ respectively.
\item Otherwise (\emph{i.e.,} if $L_i>r_c$), wait for the nodes in
  $\mathcal{S}_i$ to wake up. When node $i_k$ wakes up $(1\le k\le
  K)$, evaluate $p=\mu_k^{\pi}(W_{k}(t),b_k)$ where $b_k=
  \max\{Z_{1},...,Z_{k}\}$. If $p=1$, then transmit to the node
  $i_{\arg\max\{Z_{1},...,Z_{k}\}}$. The delay incurred is
  $D^{\pi}(t)=W_{k}(t)$ and the progress made is $Z^{\pi}(t)=b_k$. If
  $p=0$, ask the node which makes the most progress so far to stay
  awake, put the other node to sleep and wait for further nodes to
  wake up.
\item The requirement $\mu_{K}^{\pi}(w,b)=1$ in the definition of
  $\pi$ ensures that node $i$ transmits at or before the instant the
  last node wakes up.
\end{itemize}

Since the distribution of $\{(W_{k}(t),Z_{k}) : 1\le k\le K\}$ are not
dependent on the value of $t$, the average values of $D^{\pi}(t)$ and
$Z^{\pi}(t)$ also do not depend on $t$. Hence to compute these average
values we can, without loss of generality, take $t=0$ and use
$D^{\pi}$ and $Z^{\pi}$ to simplify the notation.

Let $\Pi$
 represent the class of all SH policies.
Note that many policies are excluded from class $\Pi$.
  For instance, the policy which waits for all the
  nodes to wake up and then transmits to the one which makes least
  progress does not
  belong to the class $\Pi$. This is because for a policy in $\Pi$, transmission
  is allowed only to the  node that makes the most progress so far.
 We would like to explicitly mention two SH policies namely Max Forward (MF) and First Forward(FF):

\noindent
A node using \textbf{Max Forward} policy will wait for all the  nodes in its forwarding set to wake up and then transmit to the one which makes  most progress. We use $\pi_{MF}$ to represent this policy. For this policy, $\mu_k^{\pi_{MF}}(w,b)=1$ if and only if $k=K$. This policy obtains maximum delay and maximum progress  among all other policies in class $\Pi$.\\
A node using \textbf{First Forward} policy will always transmit to the
node in the forwarding set which wakes up first irrespective of the
progress made by it. $\pi_{FF}$ is used to represent this policy. For
this policy, $\mu_1^{\pi_{FF}}(w,b)=1$. $\pi_{FF}$ obtains minimum
delay and minimum progress among all the policies in
class $\Pi$.

\section{Problem Formulation}
\label{problem_formulation}
From here on, without loss of generality we fix $T=1$ and $r_c=1$. Let
$\mathbb{P}_K$ (where $K\ge1$) denote the probability law conditioned
on the event $\{N_i=K\}$ \emph{i.e.,}
$\mathbb{P}_K(.)=\mathbb{P}(.|N_i=K)$. Similarly we define the
conditional expectation $\mathbb{E}_K$. Define
$\gamma_{MF}=\mathbb{E}_K[Z^{\pi_{MF}}]$ and
$\gamma_{FF}=\mathbb{E}_K[Z^{\pi_{FF}}]$, average progress made by the
MF and FF policies respectively.

Our interest in this work are, at a relay node $i$ with $N_i=K$, to
minimize the average delay subject to a constraint on the average
progress achieved.  More formally the problem is,
\begin{eqnarray}
\label{SH1_prob}
\min_{ {{\pi}}\in {{{\Pi}}}}&&\mathbb{E}_K[D^{{\pi}}]\\
\mbox{s.t.}&&\mathbb{E}_K[Z^{{\pi}}]\ge \gamma\nonumber
\end{eqnarray}
where $\gamma\in\left[0,\gamma_{MF}\right]$. 

This formulation embodies the one-step tradeoff between the need to
forward the packet quickly while attempting to make substantial
progress towards the sink. The parameter $\gamma$ controls the
tradeoff. A large $\gamma$ indicates our desire to make large progress
in each step, which will come at a cost of a large one hop forwarding
delay.

To solve the problem in (\ref{SH1_prob}), we consider the following
unconstrained problem, 
\begin{eqnarray}
\label{SH2_prob}
\min_{{\pi}\in{\Pi}}&&\mathbb{E}_K[D^{\pi}]-\eta\mathbb{E}_K[Z^{\pi}]
\end{eqnarray}
Where $\eta>0$.  Let $\pi_{BF}(\eta)$ (Best Forward) be the optimal
solution for this problem.
\begin{lemma} \label{lem:conversion_to_mdp}
  For a given $\gamma$ in problem (\ref{SH1_prob}), suppose there is
  an $\eta_\gamma$ such that
  $\mathbb{E}_K[Z^{\pi_{BF}(\eta_\gamma)}]=\gamma$, then
  $\pi_{BF}(\eta_\gamma)$ is optimal for the problem in
  (\ref{SH1_prob}) as well.
\end{lemma}
\begin{proof}
 Since $\pi_{BF}(\eta_\gamma)$ is optimal for the problem in (\ref{SH2_prob}), 
\begin{eqnarray*}
  \mathbb{E}_K[D^{\pi_{BF}(\eta_\gamma)}]-\eta_\gamma\mathbb{E}_K[Z^{\pi_{BF}(\eta_\gamma)}]&\le&
  \mathbb{E}_K[D^{\pi}]- \eta_\gamma \mathbb{E}_K[Z^{\pi}] \mbox{, for all } \pi\in\Pi\\
  \mbox{ \emph{i.e.} }\mathbb{E}_K[D^{\pi_{BF}(\eta_\gamma)}]&\le&\mathbb{E}_K[D^{\pi}]-
  \eta_\gamma (\mathbb{E}_K[Z^{\pi}]-\gamma)
\end{eqnarray*}
Therefore for any $\pi$ such that $\mathbb{E}_K[Z^{\pi}]\ge \gamma$, we have
\begin{equation*}
	\mathbb{E}_K[D^{\pi_{BF}(\eta_\gamma)}]\le\mathbb{E}_K[D^{\pi}]
\end{equation*}
\end{proof}
In the subsequent sections we focus on solving the problem in (\ref{SH2_prob}).

\section{Optimal Policy for the Exact Model}
\label{optimal_policy}
To solve the problem in (\ref{SH2_prob}), we develop it in a Markov
Decision Process (MDP) framework \cite{optimalcontrol}.
$\mathcal{X}=[0,1]^2\bigcup\{\psi\}$ is the state space (recall that
$T=1$ and $r_c=1$).  $\psi$ is the terminating state.
$\mathcal{C}=\{0,1\}$ is the control space where $1$ is for
\emph{stop} and $0$ is for \emph{continue}. A small change to the
$\pi$ defined earlier in section (\ref{policy_sec}), is the inclusion
of $\psi$ in the domain of $\mu^{\pi}_k$. Let $(w_k,b_k)$ be the
state at stage $k$ where $b_k$ is the best (maximum) progress made by
the nodes waking up until stage $k$ \emph{i.e.,}
$b_k=\max\{Z_{1},...,Z_{k}\}$. Conditioned on being in state
$(w_k,b_k)$ at stage $k$, transition to the next state depends on
$w_k$ through ${U}_{k+1}$ whose \emph{p.d.f.} is $f_{U_{k+1}|W_{k}}(.|w_{k})$
(Equation (\ref{condorderstat_equn})). The other disturbance component
$Z_{{k+1}}$, is independent of the $(w_k,b_k)$. \emph{p.d.f.} of $Z_{k+1}$ is
$f_{Z}$ (Equation (\ref{distribution_equn})). We define the
conditional expectation,
\begin{equation*}
 \mathbb{E}_{(W_k=w_k)}[.]=\mathbb{E}_K[.|W_k=w_k]
\end{equation*}
\noindent
Then using expression (\ref{condorderstat_equn}) we can write,
\begin{equation}
\mathbb{E}_{(W_k=w_k)}[U_{k+1}]=\frac{1-w_k}{K-k+1}
\end{equation}
\noindent
Initial state $s_0=\textbf{0}$ and initial action $a_0=0$ always.
Therefore the next state is $s_1=(U_1,Z_1)$ and the cost incurred at
stage $0$ is $g_0(\textbf{0},0)=U_1$.  If $a_k\in\mathcal{C}$ is the
action taken at stage $1\le k\le K-1$, then the next state $s_{k+1}$
is,
\begin{equation*}
s_{k+1}=\left\{ \begin{array}{ll}
                 (w_{k}+{U}_{k+1},\max\{Z_{{k+1}},b_k\})&\mbox{  if  } a_k=0\\
                \psi&\mbox{   if  } a_k=1\end{array} \right.
\end{equation*}
and the one step  cost  function is,
\begin{equation}
\label{extcostfunc_equn}
g_k((w_k,b_k),a_k)=\left\{\begin{array}{ll}
                    {U}_{k+1}&\mbox{ if }a_k=0 \\
                    -\eta b_k&\mbox{ if }a_k=1\end{array} \right.
\end{equation}
If the state at stage $k$ is $\psi$ then $s_{k+1}=\psi$ and
$g_k(\psi,a_k)=0$ irrespective of $a_k$.  Also if $s_{K}$ is the state
of the system at the last stage, there is a cost of termination,
$g_{K}(s_K)$ given as,
\begin{equation*}
g_{K}(s_{K})=\left\{\begin{array}{ll}
                                0&\mbox{ if }s_{K} =\psi\\
                                -\eta b_{K}&\mbox{ otherwise }\end{array}\right.
\end{equation*}
The total average cost incurred with policy $\pi$ is, 
\begin{equation*}
J_{\pi}(\textbf{0})=\mathbb{E}_{K}\left[\sum_{k=0}^{K-1}g_k(s_k,\mu^{\pi}_k(s_k))+g_{K}(s_{K})\right]
\end{equation*}
The expectation in the cost function above is taken over the joint distribution of $\{({U}_k,Z_{k}):1\le k\le K\}$. Note that, 
 \begin{equation*}
J_{\pi}(\textbf{0})=\mathbb{E}_K[D^{\pi}]-\eta\mathbb{E}_K[Z^{\pi}]
 \end{equation*}
 Therefore the optimal cost is,
\begin{equation*}
J^{*}(\textbf{0})=\min_\pi J_\pi(\textbf{0})=J_{\pi_{BF}(\eta)}(\textbf{0})
\end{equation*}

Let $J_k(w,b)$ be the optimal cost to go when the system is in state
$(w,b)$ at stage $1\le k\le K$. When the stage is $K$ (\emph{i.e.,}
all the nodes have woken up), then invariably transmission has to
happen. Therefore,
\begin{eqnarray}
	\label{extbell1_equn}
	J_{K}(w,b)&=&-\eta b\nonumber\\
	&=&-\eta\max\{b,\phi_{K}(w,b)\}
\end{eqnarray}
where, we define $\phi_{K}(w,b)=0$ for all $(w,b)$. Next when there is
one more node to wake up (\emph{i.e.,} stage is $K-1$) then both actions,
$a_{K-1}=1$ and $a_{K-1}=0$ are possible. Therefore,
\begin{eqnarray*}
  {J_{K-1}(w,b)}&=&\min\left\{-\eta b,\mathbb{E}_{(W_{K-1}=w)}\left[{U}_{K}+
   J_{K}(w+{U}_{K},\max\{b,Z_{K}\})\right]\right\}\nonumber
\end{eqnarray*}
The terms in the $\min$ expression are the costs when $a_{K-1}=1$
(stop) and $a_{K-1}=0$ (continue) respectively. Using the expression
for $J_K$ in (\ref{extbell1_equn}) we obtain,
\begin{eqnarray}
\label{extbell2_equn}
{J_{K-1}(w,b)}
&=&\min\left\{-\eta b, \mathbb{E}_{(W_{K-1}=w)}\left[{U}_{K}-\eta\max\{b,Z_{K},
\phi_{K}(w+{U}_{K},\max\{b,Z_{K}\})\}\right]\right\}\nonumber\\
&=&-\eta\max\{b,\phi_{K-1}(w,b)\}
\end{eqnarray}
where,
\begin{eqnarray} 
\label{phiK_equn}
\phi_{K-1}(w,b)&=&\mathbb{E}_{(W_{K-1}=w)}\left[\max\{b,Z_{K},\phi_{K}(w+{U}_{K},
\max\{b,Z_{K}\})\}-\frac{{U}_{K}}{\eta}\right]
\end{eqnarray}
The following lemma is obtained easily.
\begin{lemma}
For every $1\le k\le K-1$, the following equations holds,
\begin{equation}
	\label{optbellk_equn}
	J_k(w,b)=-\eta\max\{b,\phi_k(w,b)\}
\end{equation}
where,
\begin{eqnarray}
	\label{phik_equn}
	\phi_{k}(w,b)&=&\mathbb{E}_{(W_k=w)}\left[\max\{b,Z_{k+1},\phi_{k+1}(w+{U}_{k+1},
	\max\{b,Z_{k+1}\})\} -\frac{{U}_{k+1}}{\eta}\right]
\end{eqnarray}
\end{lemma}
\begin{proof}
  Suppose for some $2\le k\le K-1$ equations (\ref{optbellk_equn}) and
  (\ref{phik_equn}) holds, then following similar lines which was used
  to obtain  (\ref{extbell2_equn}) and (\ref{phiK_equn})
  (just replace $K$ by $k$) we can show that 
  (\ref{optbellk_equn}) and (\ref{phik_equn}) holds for $k-1$ as well.
  Since we have already shown that these equations hold for $k=K-1$,
  from induction argument we can conclude that it holds for every
  $1\le k\le K-1$.
\end{proof}
\noindent
The structure of the optimal policy is given in the following corollary.

\begin{corollary}
  The optimal policy $\pi_{BF}(\eta)$ is of the following form,
\begin{equation}
\label{optpolicy_equn}
\mu^{\pi_{BF}(\eta)}_k(w,b)=\left\{\begin{array}{ll}
                               	1&\mbox{ if }b\ge \phi_k(w,b)\\
                               	0&\mbox{ otherwise }	\end{array}\right.
\end{equation}
for $1\le k\le K$. Where $\phi_K(w,b)=0$ for all $(w,b)\in\mathcal{S}$ and for $1\le k\le K-1$, $\phi_k(w,b)$ is given in equation (\ref{phik_equn}).
\hfill $\blacksquare$
\end{corollary}

\noindent
\emph{Remarks:} The optimal policy requires threshold functions $\{\phi_k\}$ which are
computionally intensive. For our later numerical work in Section
(\ref{simulation_results}), we discretize the state space into
${10}^4$ equally spaced points and use the approximate values of the
functions $\phi_k, 1\le k\le K-1$ at these discrete points.

\section{Optimal Policy for a Simplified Model}
\label{suboptimal_policy}
\begin{figure}[ht]
\centering
\subfigure[]{
\includegraphics[scale=0.3]{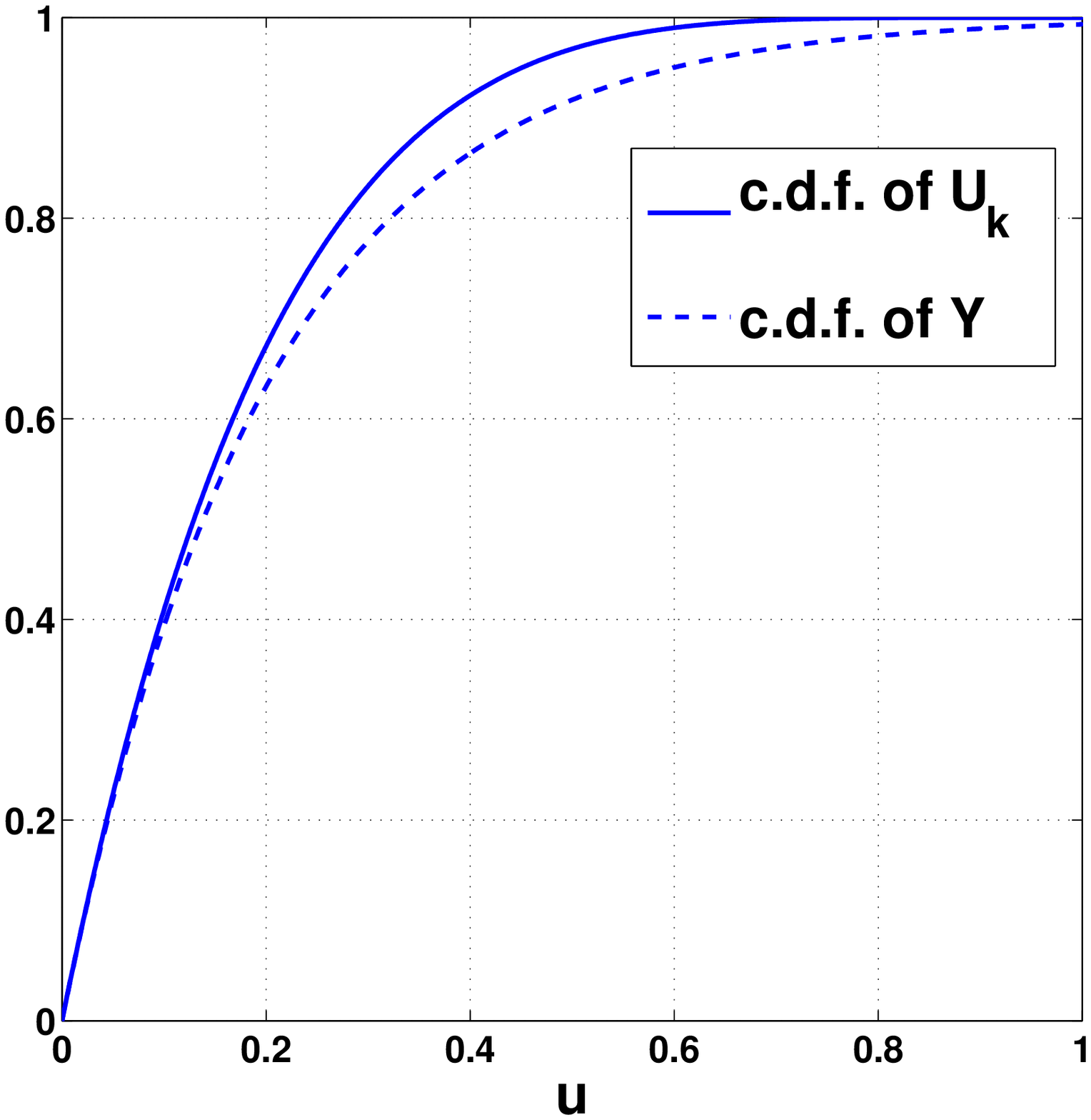}
\label{cdf1}
}
\subfigure[]{
\includegraphics[scale=0.3]{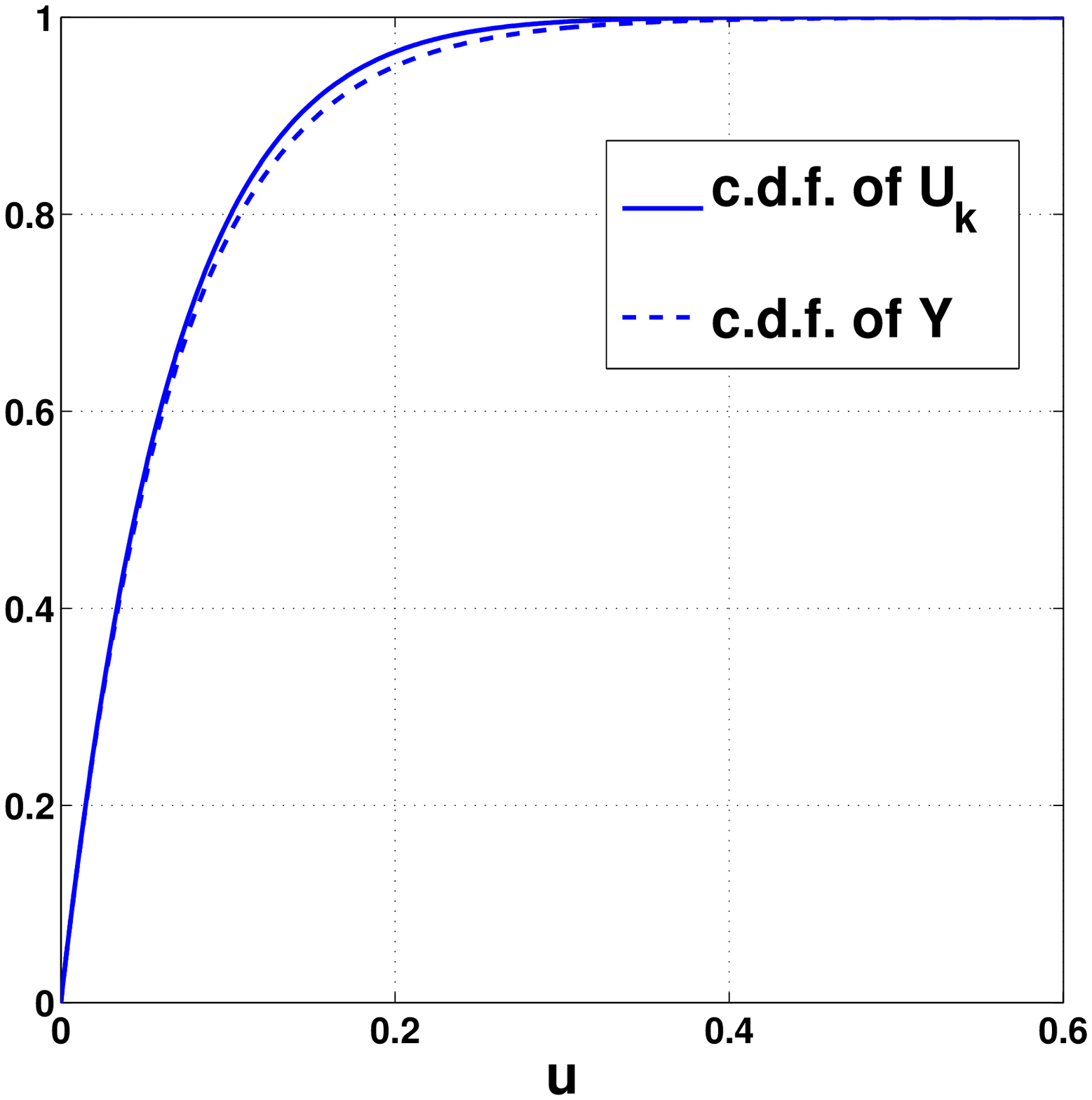}
\label{cdf2}
}
\caption{The \emph{c.d.f.}'s  $F_{U_k}$ and $F_Y$ where $Y\sim Exponential(K)$ are plotted for \subref{cdf1} $K=5$ and \subref{cdf2} $K=15$.\label{cdf_figu}}
\end{figure}

The random variables $\{U_k:1\le k\le K\}$ are identically distributed
\cite[Chapter 2]{orderstatistics} (but not independent). Their common
\emph{c.d.f.} is $F_{U_k}(u)=1-(1-u)^{K}$.  From Fig.~\ref{cdf_figu} we
observe that the \emph{c.d.f.} of $\{U_k:1\le k\le K\}$ is close to that of
the \emph{c.d.f.} of an exponential random variable of parameter $K$ and the
approximation becomes better for large values of $K$. This motivates
us to consider a \emph{simplified model} where $\{U_k:1\le k\le K\}$
are distributed as \emph{Exponential(K)}. Further in our simplified
model we take these random variables to be independent.

For the simplified model, the cost function (similar to (\ref{extcostfunc_equn})) when the system is in state $(w,b)$ at stage $1\le k\le K-1$ is,
\begin{equation}
\label{expcostfunc_equn}
g_k((w_k,b_k),a_k)=\left\{\begin{array}{ll}
                    {U}_{k+1}&\mbox{ if }a_k=0 \\
                    -\eta b_k&\mbox{ if }a_k=1\end{array} \right.
\end{equation}
We observe that due to the \emph{i.i.d.} inter-wake time assumption the cost
function is not dependent on the value of $w_k$. Also we need not
consider conditioning on $W_k=w_k$ unlike in the previous section
since the \emph{p.d.f.} of $U_{k+1}$ does not depend on $w_k$.  Hence, the
optimal policy for this model is going to be independent of $w_k$ for
each $k$.  So we simplify the state space by ignoring the values of
$w_k$ for each $k$, \emph{i.e.,} the state space is
$\mathcal{\bar{X}}=[0,1]\bigcup\{\psi\}$. Control space $\mathcal{C}$
and the other disturbance component ${Z_{k}}$ remain the same.  Since
the state space is different, we make a small change to the definition
of policy $\pi$ by allowing
$\mu_k^{\pi}:\mathcal{\bar{X}}\rightarrow\mathcal{C}$.  The state
transition and cost functions remain same as in the previous section
with $(w,b)$ replaced by $b$. Let $\pi_{SF}(\eta)$ represent the
optimal policy for this model.

Let $J_k(b)$ be the optimal cost to go at stage $k$ when the state is
$b$. Then, for all $b \in [0,1]$,
\begin{equation}
	\label{expbell1_equn}
	J_{K}(b)=-\eta b	
\end{equation}
Next when the stage is $K-1$, for $b \in [0,1]$, 
\begin{eqnarray}
\label{expbell2_equn}
J_{K-1}(b)&=&\min\{-\eta b,\mathbb{E}_K\left[{U}_{K}+J_{K}(\max\{b,Z_{K}\}) \right]\}\nonumber\\
&=&\min\{-\eta b, \mathbb{E}_K\left[{U}_{K}-\eta\max\{b,Z_{K}\}\right]\}\nonumber\\
&=&-\eta\max\{b,\beta_1(b)\}
\end{eqnarray}
where $\beta_1$ is a function, which for $b\in[0,1]$ is given by, 
\begin{eqnarray}
	\label{beta1_equn}
 	\beta_1(b)&=&\mathbb{E}_K[\max\{b,Z_K\}]-\frac{\mathbb{E}_K[{U}_K]}{\eta}\nonumber\\
&=&\mathbb{E}_K[\max\{b,Z\}]-\frac{1}{\eta K}
\end{eqnarray}	
Here we have made use of the fact that
$\mathbb{E}_K[{U}_{K}]=\frac{1}{K}$ and $Z_{K}\sim Z$. The \emph{p.d.f.} of
$Z$ is given in (\ref{distribution_equn}).  Evidently, at
stage $K-1$, the optimal action is to stop and transmit the packet if
$b \ge \beta_1(b)$ and to continue otherwise.
The following results about $\beta_1(b)$ can easily be obtained, the proof of which we provide in Appendix.
\begin{lemma}
  \label{lem:beta_properties}
  \begin{enumerate}
  \item $\beta_1$ is continuous, increasing and convex in $b$.
  \item If $\beta_1(0)<0$, then $\beta_1(b)<b$ for all $b\in[0,1]$.
  \item If $\beta_1(0)\ge 0$, then there is a unique $\alpha_{\eta}$
    such that $\beta_1(\alpha_{\eta})=\alpha_{\eta}$.
  \item If $\beta_1(0)\ge0$, then $\beta_1(b)<b$ for
    $b\in(\alpha_\eta,1]$ and $\beta_1(b)>b$ for
    $b\in[0,\alpha_\eta)$.
  \end{enumerate}
\hfill $\blacksquare$
\end{lemma}
If $\beta_1(0)<0$, then define $\alpha_\eta=0$. Otherwise
$\alpha_\eta$ is defined by $\beta_1(\alpha_\eta)=\alpha_\eta$. Then
\begin{equation*}
\mu^{\pi_{SF}(\eta)}_{K-1}(b)=\left\{\begin{array}{ll}
                                    	1 \mbox{, if } b\ge\alpha_\eta\\
                                    	0 \mbox{, otherwise}\end{array}\right.
\end{equation*}
We proceed  to evaluate $J_{K-2}$.
\begin{eqnarray}
	\label{expbell3_equn}
	J_{K-2}(b)
	&=&\min\{-\eta b,\mathbb{E}_K[U_{K-1}+J_{K-1}(\max\{b,Z_{K-1}\})]\}\nonumber\\
	&=&\min\{-\eta b,\mathbb{E}_K[U_{K-1}-\eta\max\{b,Z_{K-1},
	\beta_1(\max\{b,Z_{K-1}\})\}]\}\nonumber\\
	&=&-\eta\max\{b,\beta_2(b)\}
\end{eqnarray}
where,
\begin{eqnarray}
	\label{beta2_equn}
	\beta_2(b)&=&\mathbb{E}_K[\max\{b,Z,\beta_1(\max\{b,Z\})\}]-\frac{1}{\eta K}
\end{eqnarray}
\begin{lemma}
\label{beta2_lem}
	$\beta_2(b)\ge\beta_1(b)$ for any $b\in[0,1]$. In particular, if $b\ge\alpha_\eta$ then $\beta_2(b)=\beta_1(b)$.
\end{lemma}
\begin{proof}
  The first part follows easily because $\mathbb{E}_K[\max\{b,Z\}]$
  $\le\mathbb{E}_K[\max\{b,Z,\beta_1(\max\{b,Z\})\}]$. 
  Next, if
  $b\ge\alpha_\eta$ then from Lemma~\ref{lem:beta_properties}, 
  $\max\{b,Z\}\ge\beta_1(\max\{b,Z\})$, so that $\max\{b,Z,\beta_1(\max\{b,Z\})\}=\max\{b,Z\}$. Therefore,
	\begin{eqnarray*}
		\beta_2(b)&=&\mathbb{E}_K[\max\{b,Z\}]-\frac{1}{\eta K}
	\end{eqnarray*}
\end{proof}
\begin{lemma}
For every  $1\le k\le K-2$ the following holds,
\begin{equation}
	\label{expbellk_equn}
	J_{k}(b)=-\eta\max\{b,\beta_{K-k}(b)\}
\end{equation}
where, 
\begin{eqnarray}
\label{expbetak_equn}
\beta_{K-k}(b)&=&\mathbb{E}_K[\max\{b,Z,\beta_{K-(k+1)}(\max\{b,Z\})\}]
- \frac{1}{\eta K}\nonumber
\end{eqnarray}
and has the property, $\beta_{K-k}(b)\ge\beta_{K-(k+1)}(b)$ for any $b\in[0,1]$. In particular,
if $b\ge\alpha_{\eta}$ then $\beta_{K-k}(b)=\beta_1(b)$.
\end{lemma}
\begin{proof}
Proof is along the lines  used to obtain Equations (\ref{expbell3_equn}), (\ref{beta2_equn}) and Lemma~\ref{beta2_lem}.
\end{proof}
\vspace{2mm}
\begin{corollary}
The  policy $\pi_{SF}(\eta)$ is of the following form,\\ $\mu^{\pi_{SF}(\eta)}_{K}(b)=1$ and
\begin{equation}
	\label{exppolicy_equn}
	\mu^{\pi_{SF}(\eta)}_k(b)=\left\{\begin{array}{ll}
	                               	1&\mbox{ if }b\ge \alpha_\eta\\
	                               	0&\mbox{ otherwise }	\end{array}\right.
\end{equation}
for $1\le k\le K-1$. 
\hfill $\blacksquare$
\end{corollary}

\noindent
\emph{Remarks:} The policy is a simple one-step-look-ahead rule where
at each $k$ $(1\le k\le K-1)$ the policy compares the cost of stopping
at $k$ ($C_s=-\eta b$) with the cost of continuing for one more step
and then stopping at $k+1$
$(C_c=\frac{1}{K}-\eta\mathbb{E}_K[\max\{b,Z\}])$. The policy is to stop
if $C_s \le C_c$ (simplification yields, stop if $b\ge\alpha_\eta$),
continue otherwise.  The policy is to transmit to the first node which
makes a progress of more than $\alpha_\eta$. If all the nodes, make
progress of less than $\alpha_\eta$ then transmit to the node whose
progress is maximum at the instant the last node wakes up.
%
%
%
%
\section{Analytical Results}
\label{analytical_results}
\begin{figure}[ht]
\centering
\includegraphics[scale=0.5]{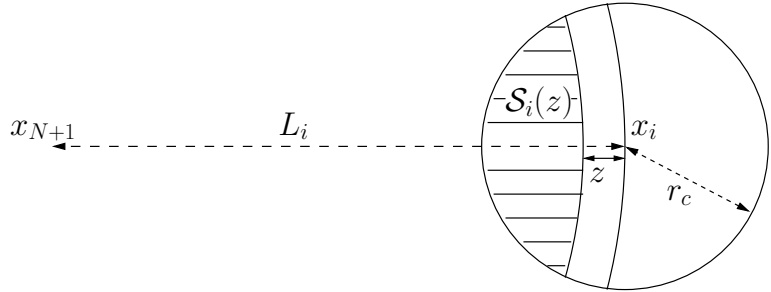}
\caption{$x_i$ and $x_{N+1}$ are the locations of node $i$ and sink
  respectively. $L_i$ is the distance between them, $r_c$ is the
  communication radius.  $\mathcal{S}_i(z)$ is the set of all points that
  are within the communication radius of node $i$ and are closer to sink
  than $i$ by atleast $z$. $\mathcal{S}_i(z)$ is the shaded region in the figure.
  \label{forwardsetoverloaded_figu}}
\end{figure}

In this section we apply the policy $\pi_{SF}(\eta)$ obtained from the
simplified model to the actual model and  obtain  expressions for 
average progress and average delay incurred by node $i$.  First we need 
some more notation. We abuse the notation $\mathcal{S}_i$ by allowing 
 $\mathcal{S}_i(z)=\{y:|y-x_i|\le r_c, |x_{N+1}-y|<L_i-z\}$. $\mathcal{S}_i(z)$ 
is the set of points that are closer to  the sink than $x_i$ by atleast $z\in[0,1]$ (see Fig. \ref{forwardsetoverloaded_figu}).  When $z=0$, we simply use  $\mathcal{S}_i$ instead of $\mathcal{S}_i(0)$. Let $p_z=\frac{|\mathcal{S}_i(z)|}{|\mathcal{S}_i|}$, where $|\mathcal{S}_i(z)|$ denotes the area of the region $\mathcal{S}_i(z)$. $p_z$ is the conditional probability that a node falls in the region $\mathcal{S}_i(z)$ conditioned on the event that the node belongs to $\mathcal{S}_i$.
\subsection{Average values for $\pi_{SF}(\eta)$}
When using policy $\pi_{SF}(\eta)$, node $i$ transmits to the first node which makes a progress of more than $\alpha_\eta$. If there are $k\ge1$ nodes in the  region $\mathcal{S}_i(\alpha_\eta)$, since the wake time of each of these is uniform on $[0,1]$ and independent of each other, the average time until the first one wakes up is $\frac{1}{k+1}$. If the region   $\mathcal{S}_i(\alpha_\eta)$ is empty, then node $i$ will wait for all the nodes in $\mathcal{S}_i$ to wake up and then transmit to the one which makes the maximum progress. In this case the average delay is $\frac{K}{K+1}$. Therefore,
 \begin{eqnarray}
        \label{expdelay1_equn}
 	\mathbb{E}_K[D^{\pi_{SF}(\eta)}]&=& \sum_{k=1}^{K}{ K \choose k }p_{\alpha_\eta}^k(1-p_{\alpha_\eta})^{K-k}\frac{1}{k+1}
	 + {(1-p_{\alpha_\eta})}^{K}\frac{K}{K+1}
 \end{eqnarray}
The expression for average progress can be written as,
\begin{eqnarray}
	\mathbb{E}_K[Z^{\pi_{SF}(\eta)}]&=&\int_0^1\mathbb{P}_K(Z^{\pi_{SF}(\eta)}>z)dz
\end{eqnarray}
When $z\in[0,\alpha_\eta]$ then  the event $(Z^{\pi_{SF}(\eta)}>z,N_i=K)$ is same as the event that at least one node makes a progress of more than $z$. 
Therefore for $z\in[0,\alpha_\eta]$,
\begin{equation}
\mathbb{P}_K(Z^{\pi_{SF}(\eta)}>z)=1-{(1-p_z)}^K
\end{equation}
When $z\in(\alpha_\eta,1]$ then the event $(Z^{\pi_{SF}(\eta)}>z,N_i=K)$ is the same as the event that the region $\mathcal{S}_i(\alpha_\eta)$ is non-empty and the node to wake up first in this region makes a progress of more than $z$, the probabilty of which is $\frac{p_z}{p_{\alpha_\eta}}$.
Therefore for $z\in(\alpha_\eta,1]$,
\begin{equation}
\mathbb{P}_K(Z^{\pi_{SF}(\eta)}>z)=\left(1-{(1-p_{\alpha_\eta})}^K\right)\frac{p_z}{p_{\alpha_\eta}}
\end{equation}
\subsection{Average Values for $\pi_{FF}$}
The policy $\pi_{FF}$ (First Forward) transmits to the node in the region $\mathcal{S}_i$ which wakes up first, irrespective of the progress made by it. Therefore,
\begin{eqnarray}
	\mathbb{E}_K[D^{\pi_{FF}}]&=& \frac{1}{K+1}
\end{eqnarray}
Average progress is,
\begin{eqnarray}
	\mathbb{E}_K[Z^{\pi_{FF}}]&=&\int_0^1\mathbb{P}_K(Z^{\pi_{FF}}>z)dz\nonumber\\
	&=&\int_0^1 p_z dz
\end{eqnarray}
\subsection{Average Values for $\pi_{MF}$}
The policy $\pi_{MF}$ (Max Forward) always waits for all the nodes to wake up and then transmits to the node which makes the maximum progress. Therefore,
\begin{eqnarray}
	\mathbb{E}_K[D^{\pi_{MF}}]&=& \frac{K}{K+1}
\end{eqnarray}
Average progress is given by, 
\begin{eqnarray}
	\mathbb{E}_K[Z^{\pi_{MF}}]&=&\int_0^1\mathbb{P}_K(Z^{\pi_{MF}}>z)dz\nonumber\\
	&=&\int_0^1 \left(1-{(1-p_z)}^K\right) dz
\end{eqnarray}
%
%
%
\section{Simulation Results}
\label{simulation_results}
\subsection{One Hop Performance}
\begin{figure*}[t]
  \centering
\subfigure[]{
\includegraphics[scale=0.26]{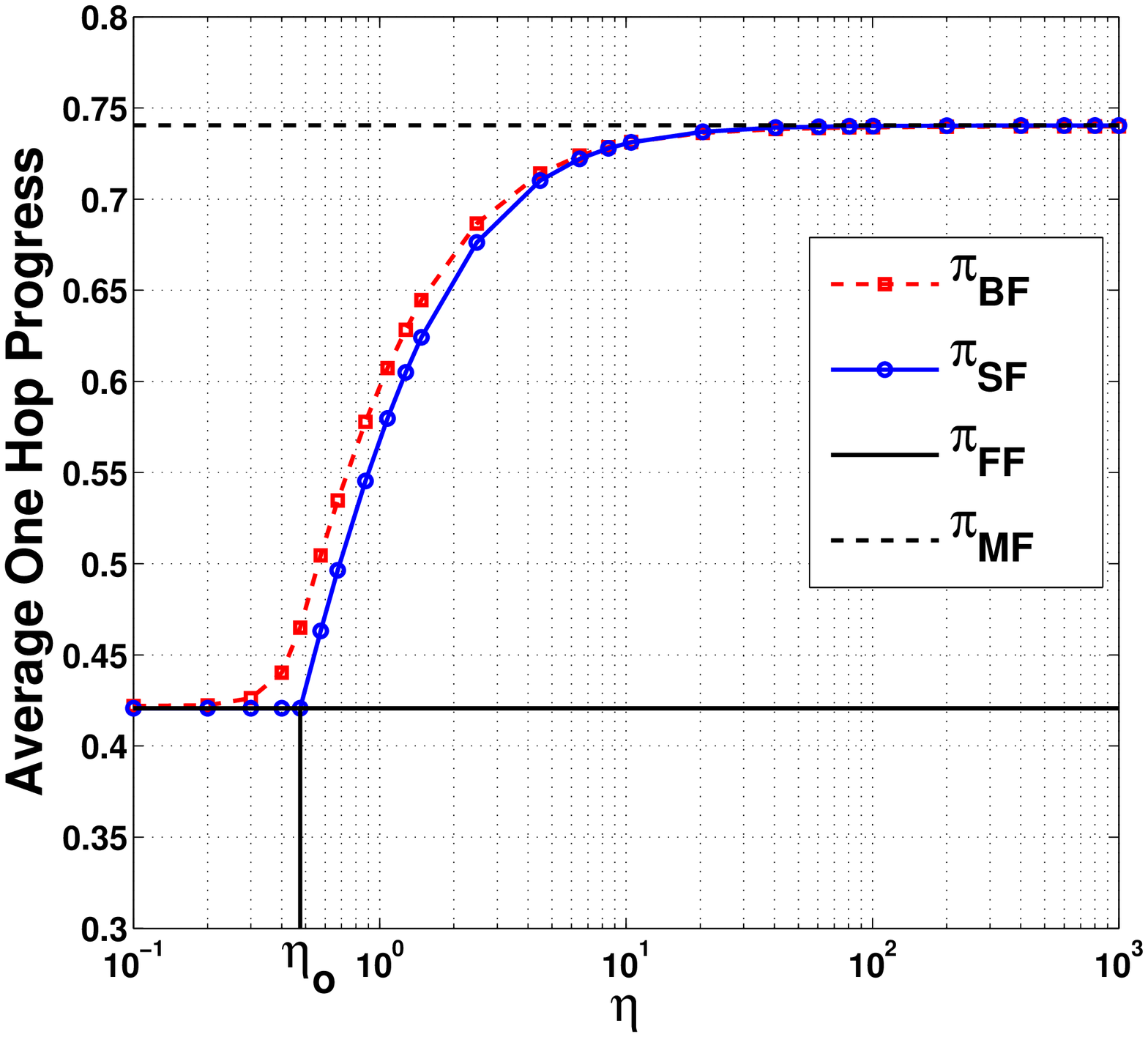}
        \label{progress_figu}
}
\subfigure[]{
\includegraphics[scale=0.26]{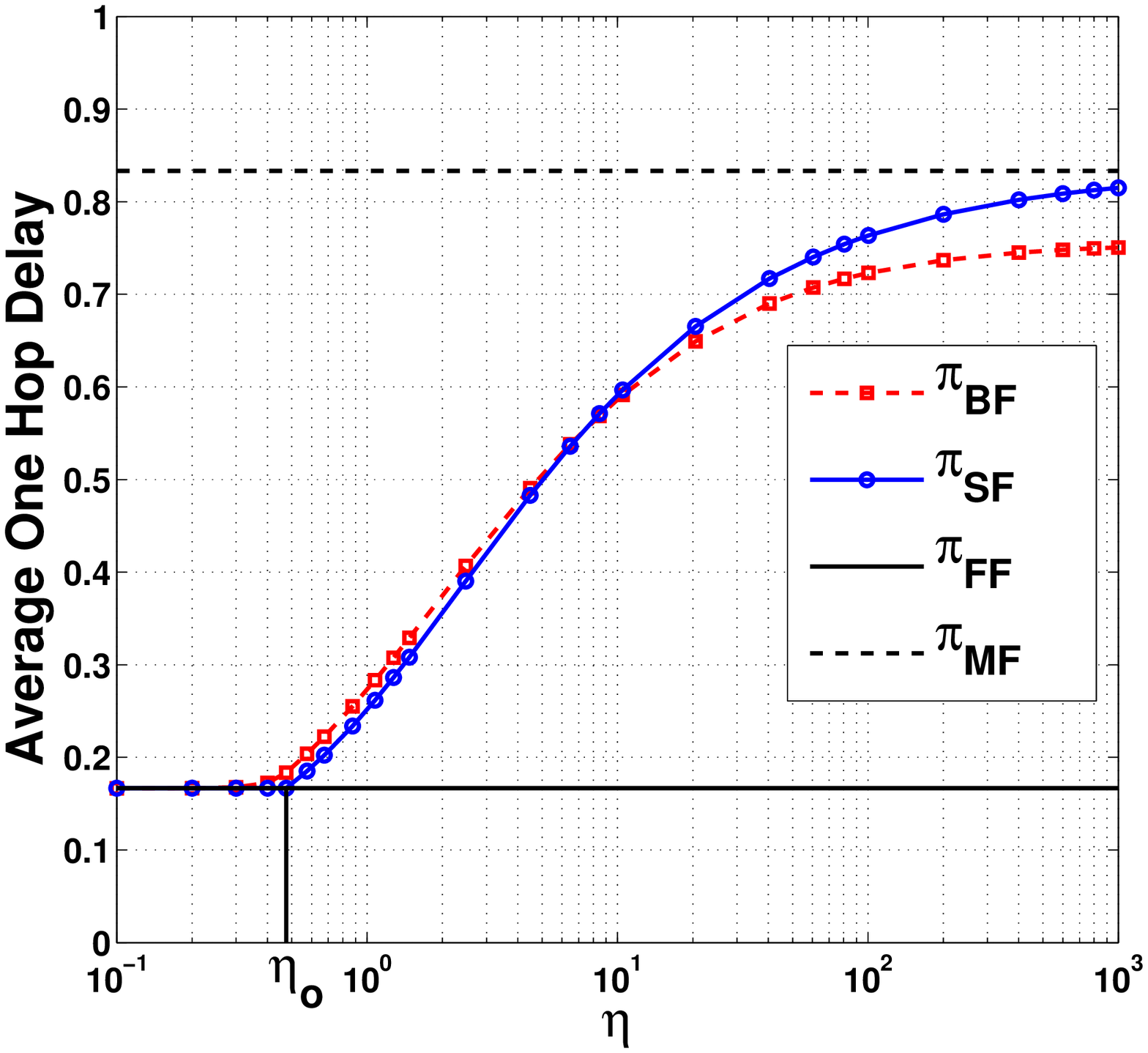}
        \label{delay_figu} 
}
\subfigure[]{
\includegraphics[scale=0.26]{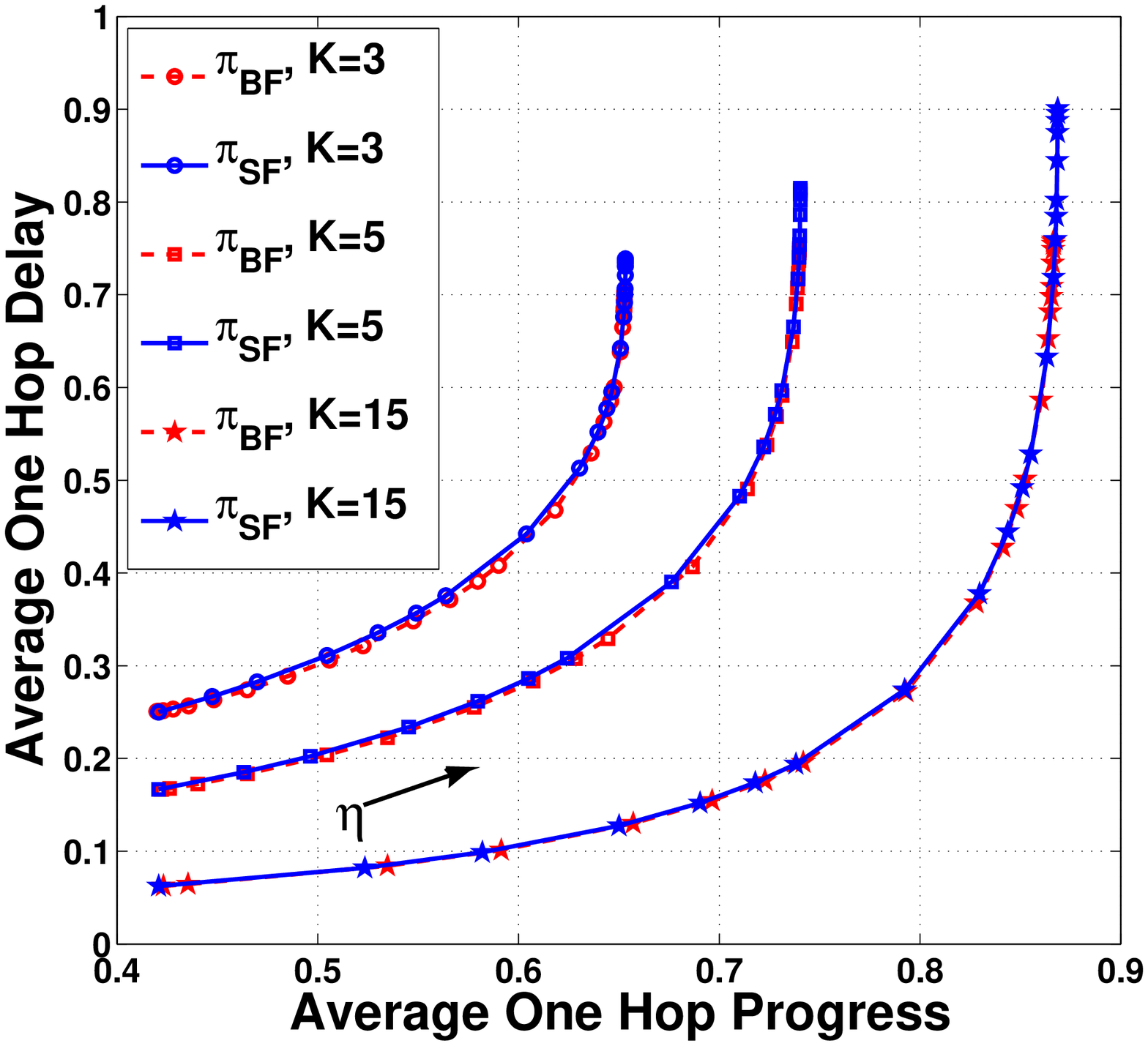}
\label{progdelay_figu}
}
\caption{One Hop Performance: \subref{progress_figu}: Average one hop progress 
  as a function of $\eta$ for  various policies. The plots are for
  $L_i=10$ and $K=5$. Maximum and minimum progress are achieved by
  $\pi_{MF}$ and $\pi_{FF}$ respectively. \subref{delay_figu}: Average one hop delay
  as a function of $\eta$ for  various
  policies. The plots are for $L_i=10$ and $K=5$. Maximum and minimum
  delay are achieved by $\pi_{MF}$ and $\pi_{FF}$. \subref{progdelay_figu}: Average one
  hop delay vs. the corresponding average one hop progress for the class of policies $\pi_{BF}$ and $\pi_{SF}$ are
  plotted for $K=3, 5$ and $15$. The  parameter $\eta$ controls the delay-progress trade-off. Each point
  on the curve corresponds to a different value of $\eta$ which
  increases along the direction shown.}
\end{figure*}

We apply the policies $\pi_{BF}(\eta)$ and $\pi_{SF}(\eta)$ to the
actual model and obtain average progress and average one hop delay for
$L_i=10$ and $K=5$. Expressions for the average values for policies
$\pi_{SF}(\eta)$, $\pi_{FF}$ and $\pi_{MF}$ were obtained in Section~\ref{analytical_results}.
 Since it is
difficult to obtain similar analytical expressions for policy
$\pi_{BF}(\eta)$, we have performed simulations to obtain these
values.  In Figs.~\ref{progress_figu} and \ref{delay_figu} we plot
the average values as a function of $\eta$.  The minimum and maximum
values of average delay and progress are achieved by $\pi_{FF}$ and
$\pi_{MF}$ respectively. From the figures we can observe that for
values of $\eta$ less than $\eta_o=\frac{1}{\mathbb{E}_{K}[Z]K}$ the
performance of $\pi_{SF}(\eta)$ is same as $\pi_{FF}$. This is because
for $\eta$ less than $\eta_o$, we have $\beta_1(0)<0$, and therefore
the threshold used is $\alpha_\eta=0$ which is same as that used by
$\pi_{FF}$.

By using a large value of $\eta$, a node will value progress more and
will end up waiting for better nodes to wake up thus incurring a large
delay as well. Hence, delay and progress for both the policies
($\pi_{BF}$ and $\pi_{SF}$) are increasing with $\eta$. We can
conclude from Lemma~\ref{lem:conversion_to_mdp}, that for each policy,
BF or SF, and a given $\eta$, the corresponding delay value is the
minimum that can be obtained using that policy, subject to a
constraint on progress equal to the progress value obtained for that
$\eta$.  These corresponding average delay vs.\ average progress values are shown in
Fig.~\ref{progdelay_figu}, for $K=3, 5 $ and $15$.  Each point on the
curve for each $K$ corresponds to a different value of $\eta$, which
increases along the curves as shown.  We see that the performance of
the $SF$ policy is close to that of the optimal BF policy, even for
small values of $K$. The way $\eta$ serves to trade-off one hop
progress and delay is clearly shown by these curves.

\subsection{End to End Performance} \label{end_subsec} 

Although our policies have been developed for one-hop optimality, it
is interesting to study their end-to-end performance if they were
used, heuristically, at each hop.  We compare the end-to-end
performance of our policy with the work of Kim et
al.~\cite{kim-etal09optimal-anycast} who have developed end-to-end
delay optimal geographical forwarding in a setting similar to ours. We
first give a brief description of their work. They minimize, for a
given network, the average delay from any node to the sink when each
node $i$ wakes up asynchronously with rate $r_i$. They show that
periodic wake up patterns obtain minimum delay among all sleep-wake
patterns with the same rate. A relay node with a packet to forward,
transmits a sequence of beacon-ID signals. They propose an algorithm
called LOCAL-OPT \cite{kim-etal08tech-report} which yields, for each
neighbor $j$ of node $i$, an integer $h_j^{(i)}$ such that if $j$
wakes up and listens to the $h-th$ beacon signal from node $i$ and if
$h \le h_j^{(i)}$, then $j$ will send an ACK to receive the packet
from $i$. Otherwise (if $h > h_j^{(i)}$) $j$ will go back to sleep. A
\emph{configuration phase} is required to run the LOCAL-OPT algorithm.

As before, we fix $r_c=1$ and $T=1$~\emph{sec}. Each node wakes up
periodically with rate $\frac{1}{T}$ but asynchronously.  To make a
fair comparision with the work of Kim et al.\ we introduce beacon-ID
signals of duration $t_I=5$ \emph{msec} and packet transmission duration of
$t_D=30$ \emph{msec}. We fix a network by placing $N$ nodes randomly in $[0,L]^2$ where $L=10$. $N$ is sampled from
\emph{Poisson($\lambda L^2$)} where $\lambda=5$.  Additional source
and sink nodes are placed at locations $(0,0)$ and $(L,L)$
respectively. Further we have considered a network where the
forwarding set of each node is non-empty.  The wake times of the
nodes are sampled independently from \emph{Uniform([0,1])}. Description of the policies that we have implemented is given below.

\noindent
\emph{$\pi_{SF}$}: We fix $\gamma$ as a network parameter.  Each relay
node chooses an appropriate $\eta$ (in other words, chooses an appropriate threshold 
$\alpha_\eta$) such that the average one hop progress made 
using the policy $\pi_{SF}(\eta)$ is equal to $\gamma$. Note 
that ${\eta}$  depends on node $i$ (\emph{i.e.,} on the values
of $L_i$ and $K$). At a relay node $i$ if $\gamma$ is less (greater) than 
the average progress made by $\pi_{FF}$ ($\pi_{MF}$) then we allow node $i$ to 
use $\pi_{FF}$ ($\pi_{MF}$) to forward. When a node $j$ wakes up and if it hears a beacon signal
from $i$, it waits for the ID signal and then sends an ACK signal containing its location information. 
If the progress made by $j$ is more than the threshold, then $i$ forwards
the packet to $j$ (packet duration is $t_D=30$~\emph{msec}). If the progress
made by $j$ is less than the threshold, then $i$ asks $j$ to stay
awake if its progress is the maximum among all the nodes that have
woken up thus far, otherwise $i$ asks $j$ to return to sleep. If more
than one node wakes up during the same beacon signal, then contentions
are resolved by selecting the one which makes the most progress among
them. In the simulation, this happens instantly (as also for the Kim et al. algorithm that we compare with); in practice this will require a splitting algorithm; see, for example, \cite[Chapter 4.3]{bertsekas-gallager87data-networks}. We assume that within $t_I=5$ \emph{msec} all these transactions
(beacon signal, ID, ACK and contention resolution if any) are over.
$\pi_{FF}$ and $\pi_{MF}$ can be thought of as special cases of
$\pi_{SF}$ with thresholds of $0$ and $1$ respectively.

\noindent
\emph{$\hat{\pi}_{SF}$}: This is same as $\pi_{SF}$ except that here a
relay node does not know $K$, but \emph{estimates} its value as
$\lfloor\lambda|\mathcal{S}_i| \rfloor$ nodes where $|\mathcal{S}_i|$ is the area of the region $\mathcal{S}_i$ (Equation~(\ref{area_equn})). If there is no eligible
node even after the $\frac{T}{t_I}-th$ beacon signal (one case when
this is possible is when the actual number of nodes $K$ is less than
$\lfloor\lambda|\mathcal{S}_i|\rfloor$ and none of the nodes make a
progress of more than the threshold) then $i$ will select one which
makes the maximum progress among all nodes.

\noindent
\emph{Kim et al.}: We run the LOCAL-OPT algorithm
\cite{kim-etal08tech-report} on the network and obtain the values
$h_j^{(i)}$ for each pair $(i,j)$ where $i$ and $j$ are neighbors. We
use these values to route from source to sink in the presence of sleep
wake cycling. Contentions, if any, are resolved (instantly, in the simulation) by selecting a node
$j$ with the highest $h_j^{(i)}$ index.

\begin{figure}
	\centering 
	\includegraphics[scale=0.5] {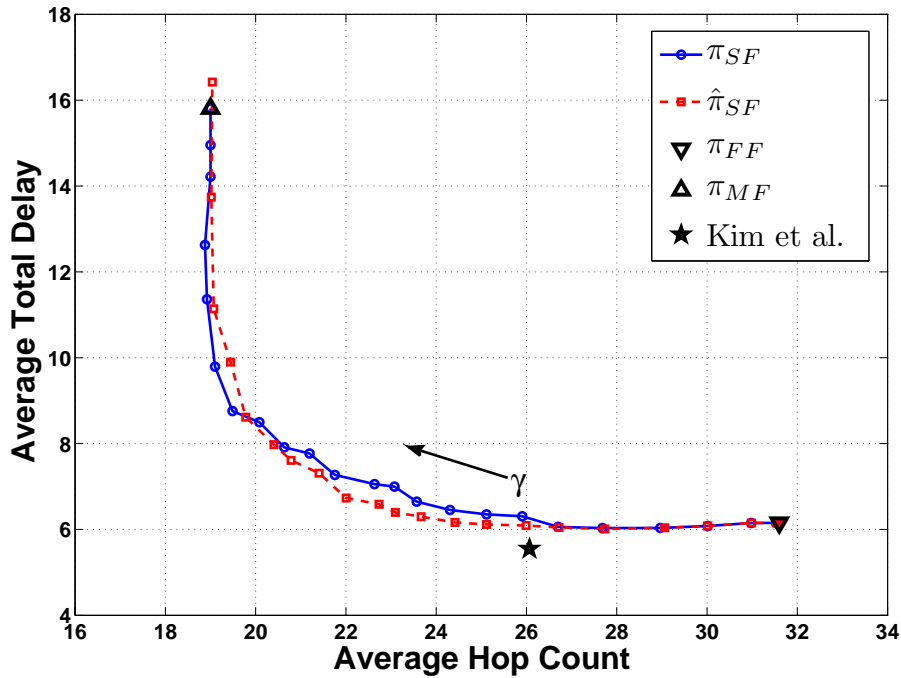}
	\caption{End-to-end performance: Plot of average end-to-end delay vs. average end-to-end hop count when the 
          one hop optimal policy for the progress constraint $\gamma$ is used at each hop.
          The  operating points of the policies $\pi_{FF}$, $\pi_{MF}$ and Kim et al.
          are also shown in the figure. Each point on the curve
          corresponds to a different value of $\gamma$ which increases
          along the direction shown. \label{endperformance_figu}}
\end{figure}

In Fig.~\ref{endperformance_figu} we plot average total delay vs.
average hop count for different policies for fixed node placement, while the averaging is over the wake
times of the nodes. Each point on the curve is obtained by averaging  over 1000 transfers of the packet from the source node to the sink.
 As expected, Kim et al.  achieves minimum average delay.
In comparision with $\pi_{FF}$, Kim et al. also achieves smaller
average hop count. Notice, however that using $\pi_{SF}$ policy and properly choosing
$\gamma$, it is possible to obtain hop count similar to that of Kim et
al., incurring only slightly higher delay. 

The advantage of $\pi_{SF}$
over Kim et al. is that there is \emph{no need for a configuration phase}.
Each relay node has to only compute a threshold that depends on the
parameter $\gamma$ which can be set as a network parameter during
deployment. A more interesting approach would be to allow the source
node to set $\gamma$ depending on the type of application. For delay
sensitive applications it is appropriate to use a smaller value of
$\gamma$ so that the delay is small, whereas, for energy constrained
applications (where the network energy needs to conserved) it is
better to use large $\gamma$ so that the number of hops (and
hence the number of transmissions) is reduced. For other applications, moderate
values of $\gamma$ can be used. $\gamma$ can be a part of the ID signal so
that it is made available to the next hop relay.

Another interesting observation from Fig.~\ref{endperformance_figu} is
that the performance of $\hat{\pi}_{SF}$ is close to that of
$\pi_{SF}$. In practice it might not be reasonable to expect a node to
know the exact number of relays in the forwarding set.
$\hat{\pi}_{SF}$ works with average number of nodes instead of the
actual number. For small values of $\gamma$ both the policies $\pi_{SF}$
and $\hat{\pi}_{SF}$, most of the time, transmit to the first node to
wake up.  Hence the performance is similar for small $\gamma$.  For
larger $\gamma$, we observe that the delay incurred by $\hat{\pi}_{SF}$ is larger.

\section{Summary and Future Work}\label{conclusion}

The problem of optimal relay selection for geographical forwarding was
formulated as one of minimizing the forwarding delay subject to a
constraint on progress.  The simple policy (SF)  of
transmitting to the first node that wakes up and makes a progress of
more than a threshold was found to be close in performance to the
optimal policy.  We then compared the end-to-end performance (average
delay and average hop count) of using SF at each relay node
enroute to the sink with that of the policy proposed by Kim et
al.~\cite{kim-etal09optimal-anycast}, which is designed to achieve
minimum average end-to-end delay. However, the delay obtained by the
 policy in \cite{kim-etal09optimal-anycast}  is only a little smaller than that obtained by the
FF policy.  Further, by using the SF policy with a
appropriate $\gamma$, performance very close to that of the policy in \cite{kim-etal09optimal-anycast}
can be obtained without the need for an initial global configuration
phase.  We note that $\pi_{SF}$ is self-configuring; each node takes
decisions based only on local information. The end-to-end performance
obtained can be tuned by the use of a single parameter $\gamma$.  For
a small $\gamma$ we obtain low end-to-end delay but the number of hops
is large and vice versa.

In this work we have assumed that each node knows the number of
neighbors in its forwarding set. We had given a heuristic policy
$\hat{\pi}_{SF}$ when the actual number of forwarding neighbors is not
known. In future work we aim to obtain optimal forwarding policies by
relaxing this assumption. Also, the use of a one-hop optimal policy
for end-to-end forwarding is  a heuristic. In future work we
propose to directly formulate the end-to-end problem and derive
optimal policies. In addition, we could also include aspects such as
the relay's link quality  in our formulation.

\bibliographystyle{IEEEtran}
\bibliography{IEEEabrv,related_work}

\begin{thebibliography}{10}
\providecommand{\url}[1]{#1}
\csname url@samestyle\endcsname
\providecommand{\newblock}{\relax}
\providecommand{\bibinfo}[2]{#2}
\providecommand{\BIBentrySTDinterwordspacing}{\spaceskip=0pt\relax}
\providecommand{\BIBentryALTinterwordstretchfactor}{4}
\providecommand{\BIBentryALTinterwordspacing}{\spaceskip=\fontdimen2\font plus
\BIBentryALTinterwordstretchfactor\fontdimen3\font minus
  \fontdimen4\font\relax}
\providecommand{\BIBforeignlanguage}[2]{{%
\expandafter\ifx\csname l@#1\endcsname\relax
\typeout{** WARNING: IEEEtran.bst: No hyphenation pattern has been}%
\typeout{** loaded for the language `#1'. Using the pattern for}%
\typeout{** the default language instead.}%
\else
\language=\csname l@#1\endcsname
\fi
#2}}
\providecommand{\BIBdecl}{\relax}
\BIBdecl

\bibitem{kim-etal09optimal-anycast}
J.~Kim, X.~Lin, and N.~Shroff, ``Optimal {A}nycast {T}echnique for
  {D}elay-{S}ensitive {E}nergy-{C}onstrained {A}synchronous {S}ensor
  {N}etworks,'' in \emph{INFOCOM 2009. The 28th Conference on Computer
  Communications. IEEE}, April 2009, pp. 612--620.

\bibitem{takagi-kleinrock84optimaltransmission}
H.~Takagi and L.~Kleinrock, ``Optimal {T}ransmission {R}anges for {R}andomly
  {D}istributed {P}acket {R}adio {T}erminals,'' \emph{Communications, IEEE
  Transactions on [legacy, pre - 1988]}, vol.~32, no.~3, pp. 246--257, 1984.

\bibitem{hou-etal86rangecontrol}
T.~C. Hou and V.~Li, ``Transmission {R}ange {C}ontrol in {M}ultihop {P}acket
  {R}adio {N}etworks,'' \emph{Communications, IEEE Transactions}, vol.~34,
  no.~1, pp. 38--44, 1986.

\bibitem{karp-etal00gpsr}
B.~Karp and H.~T. Kung, ``G{P}{S}{R}: {G}reedy {P}erimeter {S}tateless
  {R}outing for {W}ireless {N}etworks,'' in \emph{MobiCom '00: Proceedings of
  the 6th annual international conference on Mobile computing and
  networking}.\hskip 1em plus 0.5em minus 0.4em\relax New York, NY, USA: ACM
  Press, 2000, pp. 243--254.

\bibitem{kuhn-etal08algorithmicapproach}
F.~Kuhn, R.~Wattenhofer, and A.~Zollinger, ``An {A}lgorithmic {A}pproach to
  {G}eographic {R}outing in {A}d {H}oc and {S}ensor {N}etworks,''
  \emph{IEEE/ACM Trans. Netw.}, vol.~16, no.~1, pp. 51--62, 2008.

\bibitem{dulman-etal06hop-count}
S.~Dulman, M.~Rossi, P.~Havinga, and M.~Zorzi, ``On the {H}op {C}ount
  {S}tatistics for {R}andomly {D}eployed {W}ireless {S}ensor {N}etworks,''
  \emph{Int. J. Sen. Netw.}, vol.~1, no. 1/2, pp. 89--102, 2006.

\bibitem{nath-kumar08distance-hop}
S.~Nath and A.~Kumar, ``Performance {E}valuation of {D}istance {H}op
  {P}roportionality on {G}eometric {G}raph {M}odels of {D}ense {S}ensor
  {N}etworks,'' \emph{Proc. 3rd International Conference on Performance
  Evaluation Methodologies and Tools (Valuetools '08), Athens, Greece}, October
  2008.

\bibitem{mauve-hartenstein01survey}
M.~Mauve, J.~Widmer, and H.~Hartenstein, ``A {S}urvey on {P}osition-{B}ased
  {R}outing in {M}obile {A}d-{H}oc {N}etworks,'' \emph{IEEE Network}, vol.~15,
  pp. 30--39, 2001.

\bibitem{Akkaya-younis05survey}
K.~Akkaya and M.~Younis, ``A {S}urvey on {R}outing {P}rotocols for {W}ireless
  {S}ensor {N}etworks,'' \emph{Ad Hoc Networks}, vol.~3, pp. 325--349, 2005.

\bibitem{paruchuri-etal04RAW}
V.~Paruchuri, S.~Basavaraju, A.~Durresi, R.~Kannan, and S.~S. Iyengar, ``Random
  {A}synchronous {W}akeup {P}rotocol for {S}ensor {N}etworks,'' \emph{Broadband
  Networks, International Conference on}, vol.~0, pp. 710--717, 2004.

\bibitem{liu-etal07CMAC}
S.~Liu, K.~W. Fan, and P.~Sinha, ``C{M}{A}{C}: An {E}nergy {E}fficient
  {M}{A}{C} {L}ayer {P}rotocol using {C}onvergent {P}acket {F}orwarding for
  {W}ireless {S}ensor {N}etworks,'' in \emph{Sensor, Mesh and Ad Hoc
  Communications and Networks, 2007. SECON '07. 4th Annual IEEE Communications
  Society Conference on}, June 2007, pp. 11--20.

\bibitem{Zorzi-rao03geographicrandom}
M.~Zorzi, S.~Member, R.~R. Rao, and S.~Member, ``Geographic {R}andom
  {F}orwarding ({G}e{R}a{F}) for {A}d {H}oc and {S}ensor {N}etworks: {M}ultihop
  {P}erformance,'' \emph{IEEE Transactions on Mobile Computing}, vol.~2, pp.
  337--348, 2003.

\bibitem{rossi-etal08SARA}
M.~Rossi, M.~Zorzi, and R.~R. Rao, ``Statistically {A}ssisted {R}outing
  {A}lgorithms ({S}{A}{R}{A}) for {H}op {C}ount {B}ased {F}orwarding in
  {W}ireless {S}ensor {N}etworks,'' \emph{Wirel. Netw.}, vol.~14, no.~1, pp.
  55--70, 2008.

\bibitem{chaporkar-proutiere08joint-probing}
P.~Chaporkar and A.~Proutiere, ``Optimal {J}oint {P}robing and {T}ransmission
  {S}trategy for {M}aximizing {T}hroughput in {W}ireless {S}ystems,''
  \emph{Selected Areas in Communications, IEEE Journal on}, vol.~26, no.~8, pp.
  1546--1555, October 2008.

\bibitem{orderstatistics}
H.~A. David and H.~N. Nagaraja, \emph{Order Statistics (Wiley Series in
  Probability and Statistics)}.\hskip 1em plus 0.5em minus 0.4em\relax
  Wiley-Interscience, August 2003.

\bibitem{optimalcontrol}
D.~P. Bertsekas, \emph{Dynamic Programming and Optimal Control, Vol. I}.\hskip
  1em plus 0.5em minus 0.4em\relax Athena Scientific, 2005.

\bibitem{kim-etal08tech-report}
\BIBentryALTinterwordspacing
J.~Kim, X.~Lin, and N.~B. Shroff, ``Optimal {A}nycast {T}echnique for {D}elay
  {S}ensitive {E}nergy-{C}onstrained {A}synchronous {S}ensor {N}etworks,''
  2008, {T}echnical Report, Purdue University. [Online]. Available:
  \url{http://web.ics.purdue.edu/\~{}kim309/Kim08tech3.pdf}
\BIBentrySTDinterwordspacing

\bibitem{bertsekas-gallager87data-networks}
D.~Bertsekas and R.~Gallager, \emph{Data networks}.\hskip 1em plus 0.5em minus
  0.4em\relax Upper Saddle River, NJ, USA: Prentice-Hall, Inc., 1992.

\end{thebibliography}

 \appendix[Proof of Lemma \ref{lem:beta_properties}]
 \noindent
 \begin{proof}[Proof of \ref{lem:beta_properties}.1]
 Recall from Equation (\ref{beta1_equn}) that 
 \begin{eqnarray}
 \beta_1(b)&=&\mathbb{E}_K[\max\{b,Z\}]-\frac{1}{\eta K}\nonumber
 \end{eqnarray}
 Let $F_Z$ represent the $c.d.f.$ of $Z$. For  $b\in[0,1]$, the $c.d.f.$ of $\max\{b,Z\}$ is,
 \begin{equation*}
  F_{\max\{b,Z\}}(z)=\left\{\begin{array}{ll}
                     0&\mbox{ if }z<b\\
 		    F_Z(z)&\mbox{ if }z\ge b\end{array}\right.
 \end{equation*}
 \begin{eqnarray}
  \beta_1(b)&=&\int_0^1(1-F_{\max\{b,Z\}}(z))dz-\frac{1}{\eta K}\nonumber\\
 &=&b+\int_b^1(1-F_Z(z))dz-\frac{1}{\eta K}\nonumber
 \end{eqnarray}
 ${\beta_1}'(b)=F_Z(b)\ge0$  and ${\beta_1}''(b)=f_Z(b)\ge0$ implies that $\beta_1$ is continuous, increasing and convex in $b$.
 \end{proof}
 \vspace{2mm}
 \begin{proof}[Proof of \ref{lem:beta_properties}.2]
 Since $\mathbb{E}_K[\max\{b,Z\}]\le 1$, $\eta>0$ and  $K>0$, we have  $\beta_1(1)<1$. 
 Also $\beta_1$ is convex (from \emph{Lemma} \ref{lem:beta_properties}.1). Hence we can write,
 \begin{eqnarray*}
  \beta_1(b)&\le&(1-b)\beta_1(0)+b\beta_1(1)\\
 &<&b
 \end{eqnarray*}
 \end{proof}
 \vspace{2mm}
 \begin{proof}[Proof of \ref{lem:beta_properties}.3]
   Let $g(b)=b-\beta_1(b)$. Then, $g(0)\le0$ and
   $g(1)>0$ (because $\beta_1(1)<1$). Also $g(b)$ is continuous (being differentiable) on
   $[0,1]$. Hence, $\exists$ an $\alpha_{\eta}\in[0,1)$ such that
   $g(\alpha_{\eta})=0$.

 Suppose $\exists$ an ${\alpha}'_{\eta}>{\alpha_{\eta}}$ such that
 $g({\alpha}'_{\eta})=0$. Then by convexity of $\beta_1$
 (from \emph{Lemma} \ref{lem:beta_properties}.1),
 \begin{eqnarray*}
 	\beta_1({\alpha}'_{\eta})&\le&\frac{1-{\alpha}'_{\eta}}{1-{\alpha_{\eta}}}\beta_1(\alpha_{\eta})+\frac{{\alpha}'_{\eta}-{\alpha_{\eta}}}{1-{\alpha_{\eta}}}\beta_1(1)\\
 \end{eqnarray*}
 $i.e.,$ ${\beta_1}(1)\ge1$. Contradicts the fact that, $\beta_1(1)<1$.
 \end{proof}
 \vspace{2mm}
 \begin{proof}[Proof of \ref{lem:beta_properties}.4]
   Again consider $g(b)=b-\beta_1(b)$. $g(b)$ is continuous (being
   differentiable) on $[0,1]$.  Suppose $\exists$ $b\in(\alpha_\eta,1]$
   such that $\beta_1(b)>b$, then $g(b)\le0$ and $g(1)>0$. This implies that
   $\exists$ $b'$ in $[b,1)$ such that $g(b')=0$. Contradicts the
   uniqueness of $\alpha_\eta$ shown in \emph{Lemma} \ref{lem:beta_properties}.3.  Similarly
   it can be shown that $\beta_1(b)>b$ for $b\in[0,\alpha_\eta)$.
 \end{proof}
\end{document}